\documentclass[pra]{revtex4}
\usepackage{amsmath}
\usepackage{graphicx}
\begin{document}
\title{Simple Model of Splitting Instability in Swollen Membranes}
\author{Hidetsugu Sakaguchi and Satomi Maeyama}
\affiliation{Department of Applied Science for Electronics and Materials,
Interdisciplinary Graduate School of Engineering Sciences, Kyushu
University, Kasuga, Fukuoka 816-8580, Japan}
\begin{abstract}
A simple one-dimensional mechanical model is proposed for splitting instability in swollen membranes. 
The splitting instability occurs by ring constriction. The bifurcation can be both subcritical and supercritical, depending on the slenderness of spheroidal membranes. The lower critical point is estimated theoretically for the simplest circular system. Necking instability occurs in a spheroidal membrane when the twist angle in the boundary conditions is increased.  
\end{abstract}
\maketitle
\section{Introduction}
Membranes separate space into the inner and outer spaces. 
The outer space becomes an environment for the object localized in the inner space. The cellular membrane separates a living organism from the environment.   Giant vesicles are nonbiological lipid membranes. They have been studied as a model system of biological membranes. Giant vesicles take various shapes, depending on the environmental conditions such as osmotic pressure and temperature~\cite{rf:1}. Different types of shape deformation are observed when the phase separation occurs in the lipid membrane~\cite{rf:2}  
The deformation of vesicles has been studied by several authors~\cite{rf:3,rf:4,rf:5}. Elastic energy is an important factor for understanding the various shapes of vesicles. The elastic energy of the deformation of the elastic membrane is the sum of the strain energy along the surface and the bending energy~\cite{rf:6}. The former  deformation is further divided into expansion-contraction and  shear.  Concerning the problems of the deformation of vesicles, the elastic energy along the surface does not play an important role because the surface area is almost constant, and the shear strain does not appear because the lipid membrane is a fluid membrane. 
Elastic sheets are other research subjects, in which elastic energy plays an essential role. The crumpling of elastic sheets has been intensively studied because of the complex folding patterns and singular behavior they generate~\cite{rf:6,rf:7}. Concerning the problems of the crumpling of elastic sheets, only bending energy is important. The surfaces of swollen or shrinking gels also exhibit interesting patterns~\cite{rf:8,rf:9}, although swollen gels are three-dimensional materials and not a membrane. Concerning the problems of the surface deformation of gels, both bending energy and deformation energy along the surface are important.  In general, the deformation of an elastic body and instabilities such as  buckling have a long history since Euler's analysis of incipient bucking in the 18th century. The von-Karman equations are the formal basis for understanding strongly deformed elastic bodies~\cite{rf:10},

In biological membranes, cell division is an interesting and important problem from the viewpoint of large deformation. 
The self-replication of vesicles and micelles was observed to be driven by chemical energy, and these structures might be interpreted as a simple model system of a primitive cell~\cite{rf:11,rf:12}. We proposed a Ginzburg-Landau-type equation for the splitting of cellular structures~\cite{rf:13}. The self-replication of these systems occurs owing to the instability of a homogeneous system.    
However, the process of cell division of biological cells is different from such a splitting process. Cell division is well controlled by many kinds of genes and proteins~\cite{rf:14}. Several mathematical models of the cell division cycle were proposed on the basis of the kinetic equations of important genes and proteins~\cite{rf:15} In cell division, the cellular membrane strongly contracts along the cleavage furrow by the filaments of actin and myosin, and two cells are produced as a result of the splitting process. That is, the cellular membrane is locally deformed at the midplane of the cell by ring constriction. The mechanics of cytokinesis has been intensively studied in biophysics~\cite{rf:16,rf:17,rf:18}. 
Some numerical simulations of cell division have been performed~\cite{rf:19}. However, in spite of these many investigations, the dynamics of cell division is not yet completely understood. 

In this paper, we propose a very simple model of splitting. 
Three-dimensional elastic deformations are generally difficult for numerical simulation, although some software programs for surface evolutions have been developed~\cite{rf:20}. 
Our system is a very simple model, it is easy to use for numerical simulation and suitable for the qualitative understanding of splitting processes.
  Our model is based on cell division driven by the ring contraction of actin filaments, that is, the cell contracts by an external force at the midplane of the cell.  However, detailed mechanisms of the biological cell-splitting process are not incorporated in our simple model. We consider a one-layer membrane such as a balloon. The membrane shrinks owing to an elastic attractive force and expands by internal pressure. The deformation along the surface is important in this system and we neglect  bending energy for simplicity. 

We assume the rotational symmetry around the $y$-axis or a spheroidal membrane, that is, the cross section perpendicular to the $y$-axis is assumed to be a circle. 
Our system is a ring chain model of $N$ particles coupled with springs under 
pressure.   
Spheroidal membranes constrict by an external force at the midplane, which is analogous to the ring constriction in cell division.  The expansion-contraction deformation is important in this process. 
We also consider twist deformation for a spheroidal membrane. Shear deformation is important in this process. 
We discuss necking instability and bifurcation structures from the results of the numerical simulation of a one-dimensional model.  Related buckling phenomena by  twist force were studied in elastic rods or cylinders; however, the pressure effect is not incorporated in these systems~\cite{rf:21}.  

The linear elastic theory is almost complete~\cite{rf:22}, however,  nonlinearity plays an important role in the instability and the behavior after the instability. In our model, elasticity is represented by a linear chain of springs, and  nonlinearity is involved in the pressure term.

\section{Splitting Instability in a One-Dimensional Membrane Model}
We first consider a closed curve in a two-dimensional space, which separates the inner and outer spaces. The closed curve consists of $N$ particles located at ${\bf r}_i=(x_i,y_i,0)$, that is,  $N$ particles are set on a plane $z=0$ in the three-dimensional space. We assume that the closed curve shrinks by elastic force and expands by internal pressure.  To express these properties, we assume that each particle attracts  neighboring particles and expands in a direction orthogonal to the closed curve. 
To describe the dynamical behavior, the equation of motion for the particles is assumed to be 
\begin{equation}
\frac{d{\bf r}_i}{dt}=K({\bf r}_{i+1}-2{\bf r}_i+{\bf r}_{i-1})+P\frac{{\bf r}_{i+1}-{\bf r}_{i-1}}{|{\bf r}_{i+1}-{\bf r}_{i-1}|}\times {\bf e}_z,
\end{equation}
where $K$ is the spring constant, $P$ denotes the strength of the pressure-like force, and ${\bf e}_z=(0,0,1)$  is the unit vector in the $z$-direction. The second term on the right-hand side indicates that the direction of the pressure-like force is orthogonal to the closed curve. Periodic boundary conditions are assumed at $i=1$ and $i=N$, owing to the closed curve. The energy $E$ in this system is expressed as
\begin{equation}
E=\frac{1}{2}K\sum|{\bf r}_{i+1}-{\bf r}_i|^2-PV,
\end{equation}
where $V$ is the area inside the closed curve. By the long-time evolution of Eq.~(1), a stable stationary state is obtained. The stationary state is a circle, as shown in Fig.~1(a), at $K=64,N=1000$ and $P=0.006$. Figure 1(b) shows the radius $R$ of the circle as a function of $P$ at $K=64$ and $N=1000$. In the stationary state, the particles are located on the circle, and the angle between neighboring particles is $\phi=2\pi/N$. The elastic force between neighboring particles is expressed as $f=2KR\sin(\phi/2)$, and the force in the direction of the center of the circle is evaluated at $f^{\prime}=2KR\sin^2(\phi/2)$, as shown in Fig.~1(c). Since there are two neighboring particles, the total force in the direction of the center is  $4KR\sin^2(\phi/2)$, that is, a shrinking force proportional to the radius $R$ works in our simple ring-chain model. On the other hand, the pressure-like force is assumed to be $P$. The equilibrium condition of the two forces therefore yields
\begin{equation}
R=\frac{P}{4K\sin^2{\phi/2}}\sim \frac{PN^2}{4K\pi^2}. 
\end{equation}
The dashed line in Fig.~1(b) shows the line of $R=PN^2/(4K\pi^2)$ at $N=1000$ and $K=64$, which is consistent with the results of direct numerical simulations. 
\begin{figure}[t]
\begin{center}
\includegraphics[height=4.cm]{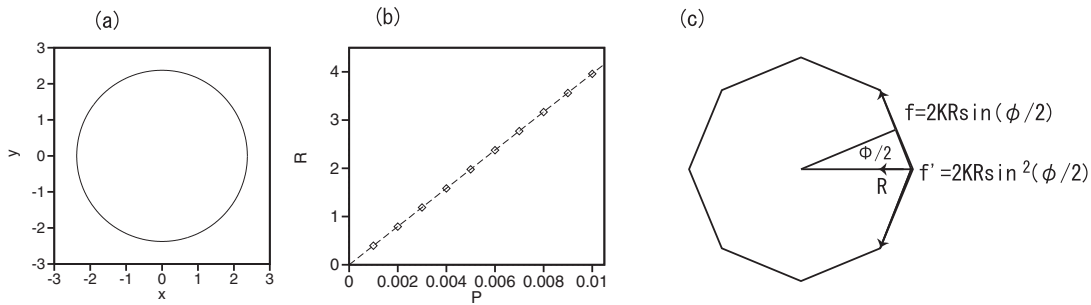}
\end{center}
\caption{(a) Equilibrium profile of $(x_i,y_i)$ by Eq.~(1) at $N=1000,K=64$, and $P=0.006$. (b) Relationship of $R$ and $P$ at $N=1000$ and $K=64$. (c) Elastic force $f$ between neighboring particles located at $(R,0)$ and $(R\cos\phi,R\sin\phi)$, and its horizontal component $f^{\prime}$. }
\label{f1}
\end{figure}
\begin{figure}[t]
\begin{center}
\includegraphics[height=4.8cm]{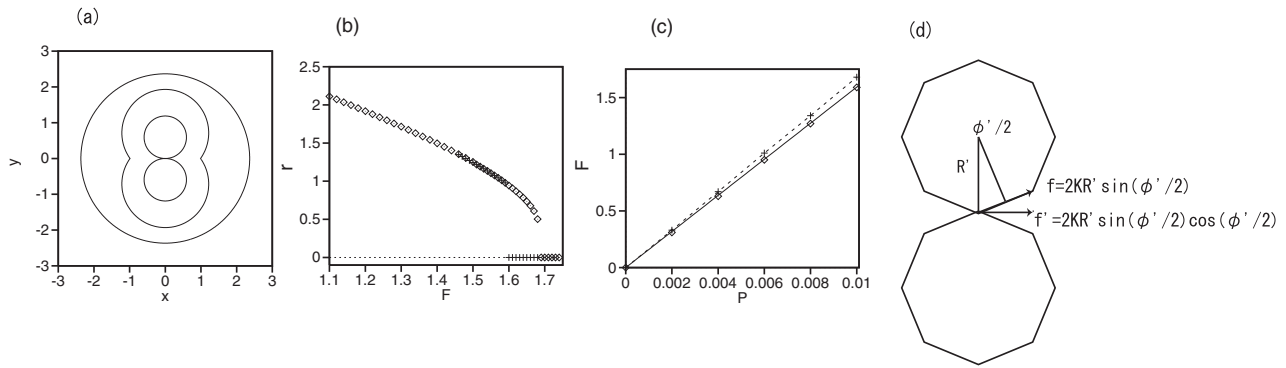}
\end{center}
\caption{(a) Equilibrium profile of $(x_i,y_i)$ obtained by Eqs.~(1) and (4) at $F=0,0.8$, and 1.6 for $N=1000,K=64$ and $P=0.006$. (b) Relationship of $r$ and $F$ at $P=0.01$, $N=1000$, and $K=64$. (c) Upper and lower critical values of $F$ as a function of $P$ for $N=1000$ and $K=64$. (d) Elastic force $f$ between neighboring particles located at $(0,-R^{\prime})$ and $(R^{\prime}\sin\phi^{\prime},-R^{\prime}\cos\phi^{\prime})$, and its horizontal component $f^{\prime}$.}
\label{f2}
\end{figure}

To constrict the cell, a force of $F$ is applied in the $-x$-direction for a particle located at an angle $\theta=0$ ($i=1$) and a force of the same strength is applied in the $x$-direction for a particle located at $\theta=\pi$ ($i=N/2+1$).  Then, the equation of motion for $i=1$ and $i=N/2+1$ is expressed as
\begin{equation}
\frac{d{\bf r}_i}{dt}=K({\bf r}_{i+1}-2{\bf r}_i+{\bf r}_{i-1})+P\frac{{\bf r}_{i+1}-{\bf r}_{i-1}}{|{\bf r}_{i+1}-{\bf r}_{i-1}|}\times {\bf e}_z\mp F{\bf e}_x,
\end{equation}
where $\mp F$ takes $-F$ for $i=1$ and $F$ for $i=N/2+1$, and ${\bf e}_x=(1,0,0)$ is the unit vector in the $x$-direction. 
The other particles obey Eq.~(1)

Figure 2(a) shows the equilibrium profiles of $(x_i,y_i)$ at $F=0,0.8$, and 1.6 for $N=1000,K=64$, and $P=0.006$. At $F=1.6$, the cell splits into two. A strong repulsive interaction is expected to work and a reconnection of neighboring particle pairs occurs in an actual splitting process, when $x_1$ and $|x_{N/2+1}|$ decrease to 0. However, such processes are not included in Eqs.~(1) and (4). Instead, $x_1$ and $x_{N/2+1}$ are assumed to be fixed at 0, when they decrease to 0. 
The two-cell state at $F=1.6$ is a result obtained  using the additional assumption.  The two cells take a circular form of radius $R^{\prime}$. $R^{\prime}$ is evaluated as $R^{\prime}\sim PN^2/(16K\pi^2)$ because $N/2$ particles are included in each circle. The transition is hysteretic. When $F$ is increased, the one-cell state becomes unstable and splits into the two-cell state.
When $F$ is decreased from a large value, the two-cell state changes into the one-cell state. Figure 2(b) shows half of the distance $r$ between the two forced particles located at $\theta=0$ and $\pi$ for $P=0.01$. The one-cell state becomes unstable at $F=1.685$ and $r$ jumps from $r=0.503$ to 0. The two-cell state becomes unstable at $F=1.595$ and $r$ jumps from 0 to $r=0.978$. The two critical values of $F$ are plotted in Fig.~2(c) as a function of $P$.

The lower critical value can be theoretically estimated. 
The elastic force between neighboring particles in the two-cell state is evaluated as $f=K2R^{\prime}\sin(\phi^{\prime}/2)$, where $\phi^{\prime}=2\pi/(N/2)$ is the angle between neighboring particles. The horizontal component of the force at a contact point is $f'=K2R^{\prime}\sin(\phi^{\prime}/2)\cos(\phi^{\prime}/2)$.  
The horizontal force in the $x$-direction for the particle located at $\theta=0$ is evaluated as $4KR^{\prime}\sin(\phi^{\prime}/2)\cos(\phi^{\prime}/2)$ because there are two neighboring particles. At the lower critical point, the horizontal force is equal to $F$, and thus,    
\begin{equation}
F_c=4KR^{\prime}\sin(\phi^{\prime}/2)\cos(\phi^{\prime}/2)\sim \frac{NP}{2\pi} \end{equation}
is satisfied. This force $F_c\sim NP/(2\pi)$ is  the critical value of remerging. The solid line in Fig.~2(c) is $NP/(2\pi)$ for $N=1000$, which is a good approximation of the lower critical value. 

\begin{figure}[t]
\begin{center}
\includegraphics[height=5.cm]{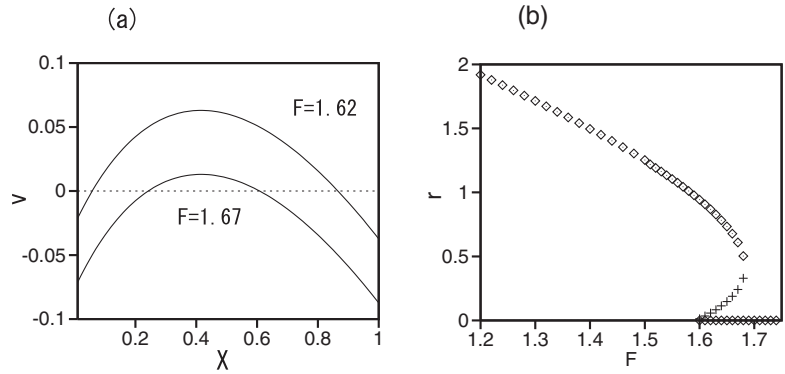}
\end{center}
\caption{(a) Relationship of $x_1=X$ and $v=dx_1/dt$ at $F=1.62$ and 1.67. (b) Stationary values $r=x_1$ as a function of $F$. The rhombi denote the stable solutions and the plus marks denote the unstable solutions.}
\label{f3}
\end{figure}
To study the bifurcation structure, we need to find an unstable solution. To find an unstable solution, we use other fixed boundary conditions that $x_1=-x_{N/2+1}=X, y_1=y_{N/2+1}=0$ for the equation of motion, Eq.~(1), and $v=dx_1/dt$ is evaluated using Eq.~(4). Figure 3(a) shows the relationship of $v$ and $X$ for $F=1.62$ and $1.67$. The intersection points of $v=v(X)$ and the horizontal line $v=0$ are stationary solutions of the equations of motion, Eq.~(1) and Eq.~(4).  There are two intersection points; however, the right one is stable and the left one is unstable. Figure 3(b) shows the stationary solutions as a function of $F$. The subcritical bifurcation structure is clearly seen in Fig.~3(b). 

Next, we consider a two-dimensional closed surface or membrane in the three-dimensional space. 
For simplicity, we assume the rotational symmetry around the $y$-axis, that is, we assume a spheroidal membrane. In other words, the cross section perpendicular to the $y$-axis is a circle for the spheroidal membrane, as shown in Fig.~4(a). The circle is assumed to be composed of $N^{\prime}$ particles 
coupled with springs of spring constant $K^{\prime}$, although it is virtual and not explicitly used in the modeling. 
In the modeling, we consider a ring chain of $N$ particles coupled with springs located in the cross section at $z=0$. 
The $i$th particle is located at ${\bf r}_i=(x_{i},y_{i},0)$ ($i=1,\cdots,N$). 
Each particle interacts with neighboring particles, and pressure expands the membrane. An additional force works in the $x$-direction, because we assume that there is a virtual circular chain of $N^{\prime}$ particles coupled with springs of spring constant $K^{\prime}$ in the cross section perpendicular to the $y$-axis. The squeezing force originating from the elastic force in the virtual circular chain of particles is expressed as $f_i^{r}=4K^{\prime}x_i\sin^2(\phi^{\prime}/2)$, where $\phi^{\prime}=2\pi/N^{\prime}$. This is because the dynamics on the closed circle in the cross-section perpendicular to the $y$-axis is similar to that explained in Figs.~1(a) and 1(c), where $x_i$ is the radius in Fig.~1(c), $K^{\prime}$ is the spring constant for the spring along the circle in the cross section, and $\phi^{\prime}=2\pi/N^{\prime}$ is the angle between nearest-neighboring particles  along the circle. A squeezing force appears owing to the attractive interaction along the virtual circular chain. This squeezing force in the $x$-direction is expressed as $-K^{\prime\prime}x_i=-4K^{\prime}\sin^2(\pi/N^{\prime})x_i$. In our numerical simulations, $N^{\prime}$ is assumed to be equal to $N$.  Even if $N^{\prime}=N$ and $K^{\prime}=K$, the elastic forces are not isotropic, because the squeezing force by the virtual circular chain expressed as $K^{\prime\prime}(-x_i,0)$ is not orthogonal to the shrinking force: $K({\bf r}_{i+1}-2{\bf r}_i+{\bf r}_{i-1})$. 
 The spherical shape shown in Fig.~1 corresponds to that in the case of $K^{\prime\prime}=0$.  By this squeezing force, the spherical shape is deformed to a spheroidal shape when $K^{\prime\prime}>0$.  
 
Furthermore, the ring constriction force is assumed to work at $y=0$. 
In a problem of the deformation of a balloon, it corresponds to a situation in which  a ring constriction force is applied at the midplane of a long balloon set in the $y$-direction. 
In our one-dimensional model in the section of $z=0$, the constriction forces $\mp F$ are applied to the particles of $i=1$ and $i=N/2+1$, as shown in Fig.~4(a).     
Then, the model equation for the particle positions is expressed as 
\begin{eqnarray}
\frac{d{\bf r}_i}{dt}&=&K({\bf r}_{i+1}-2{\bf r}_i+{\bf r}_{i-1})+P\frac{{\bf r}_{i+1}-{\bf r}_{i-1}}{|{\bf r}_{i+1}-{\bf r}_{i-1}|}\times {\bf e}_z-K^{\prime\prime}x_i{\bf e}_x,\;\;\;{\rm for}\;\; i\ne 1,N/2+1,\nonumber\\
\frac{d{\bf r}_i}{dt}&=&K({\bf r}_{i+1}-2{\bf r}_i+{\bf r}_{i-1})+P\frac{{\bf r}_{i+1}-{\bf r}_{i-1}}{|{\bf r}_{i+1}-{\bf r}_{i-1}|}\times {\bf e}_z+(-K^{\prime\prime}x_i\mp F){\bf e}_x,\;\;\;{\rm for}\;\; i=1, N/2+1.\nonumber\\
\end{eqnarray}
The model equation is almost the same as Eqs.~(1) and (4), except that the force  $-K^{\prime\prime}x_i$ in the $x$-direction is added.  Equations (1) and (4) are special cases of Eq.~(6) of $K^{\prime\prime}=0$.

Figure 4(b) shows stationary profiles of $(x_i,y_i)$  at $F=0,1$, and 2 for $N=1000,K=64,P=0.01$, and $K^{\prime\prime}=4K\sin^2(\pi/N)$ ($K^{\prime}=K$). 
At $F=0$, an elliptic curve appears. Compared with that in Fig.~1(a), a circle becomes an ellipse owing to the additional squeezing force $-K^{\prime\prime}x$. 
At $F=1$, a neck structure appears at $y=0$ owing to the constriction force at $y=0$.  At $F=2$, the ellipse is split into two ellipses. 
Figure 4(c) shows the relation of $X=x_i$ at $i=1$ as a function of $F$ for $N=1000,K=64,P=0.01$, and $K^{\prime\prime}=4K\sin^2(\pi/N)$ ($K^{\prime}=K$). $X$ decreases to zero continuously. The bifurcation is supercritical in contrast to that in the case of $K^{\prime\prime}=0$ shown in Fig.~2(b). That is, the subcritical bifurcation changes into the supercritical bifurcation, as $K^{\prime\prime}$ increases from 0 and spheroidal membranes become slender.
 
\begin{figure}[t]
\begin{center}
\includegraphics[height=5.cm]{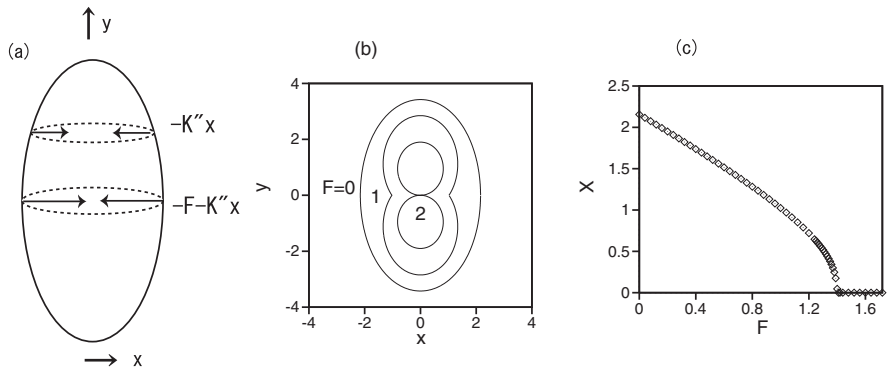}
\end{center}
\caption{(a) Spheroidal membrane. The $x$ and $y$-axes are in the horizontal and vertical directions respectively, and the $z$-axis is directed out of the paper.  
The cross sections shown by dashed curves perpendicular to the $y$-axis are circles. Centripetal forces work at every $i$ and the ring constriction force works only for $i=1$ and $N/2+1$ at $y=0$.  
(b) Equilibrium profiles of $(x_i,y_i)$ at $F=0,1$, and 2 for $N=1000,K=64,P=0.01$, and $K^{\prime\prime}=4K\sin^2(\pi/N)$. (c) $X=x_1$ as a function of the constriction force $F$.}
\label{f4}
\end{figure}

\section{Necking Instability by Twist Deformation}
We can apply the simple model of swollen membranes to a problem of twist deformation. In balloon art, various shapes are created by constructing neck structures of almost 0 width in a long and slender balloon. The neck structures are created by applying twist forces at the two ends of a balloon. 
A similar type of deformation by twisting is used in a process of making sausages, that is, the characteristic sausageshape appears by twisting the casing made from bowels.  A schematic figure for the twist of a spheroidal membrane is shown in Fig.~5(a). 

We consider a one-dimensional system of $N$ particles coupled with springs as described before. The position of the $i$th particle is expressed as ${\bf r}_i=(x_i,y_i,z_i)$. The twist deformation is set along the circumferential direction perpendicular to the $y$-axis in the two end regions.  
To make a twisted state, we first construct a spheroidal membrane using Eq.~(6) with $F=0$. At this stage, the one-dimensional chain of particles is located on $z=0$, and the position of the $i$th particle is expressed as ${\bf r}_i=(x_{i0},y_{i0},0)$.
 Then, we impose twisted boundary conditions for particles near the two ends as
\begin{eqnarray}
x_{i}&=&x_{0i}\cos\phi_0,\;z_{i}=x_{0i}\sin\phi_0,\;y_{i}=y_{i0}
,\;\;\;{\rm for}\;\; 3N/16<i<5N/16,\nonumber\\
x_{i}&=&x_{0i}\cos\phi_0,\;z_{i}=-x_{0i}\sin\phi_0,\;y_{i}=y_{i0}
,\;\;\;{\rm for}\;\; 11N/16<i<13N/16.
\end{eqnarray}
That is, the particles near the top and bottom ends are twisted by an angle $\pm\phi_0$. The other particles are assumed to obey the following equation of motion:
\begin{equation}
\frac{d{\bf r}_i}{dt}=K({\bf r}_{i+1}-2{\bf r}_i+{\bf r}_{i-1})+P\frac{{\bf r}_{i+1}-{\bf r}_{i-1}}{|{\bf r}_{i+1}-{\bf r}_{i-1}|}\times {\bf e}_t-{\bf f}_i^{r},
\end{equation}
where $-{\bf f}_i^{r}=K^{\prime\prime} (-x_i,0,-z_i)$ is the squeezing force by the virtual circular chain, ${\bf e}_t$ is a unit vector $\pm(-z_i/\sqrt{x_i^2+z_i^2},0,x_i/\sqrt{x_i^2+z_i^2})$ perpendicular to the direction of $(x_i,0,z_i)$. Here, $\pm$ respectively corresponds to particles satisfying $i\le N/4, 3N/4+1\le i$ and $N/4+1\le i\le3N/4$. Owing to this assignment, a pressure-like force  works in the outward direction perpendicular to the spheroidal membrane.  
Note that the ring constriction force $\mp F{\bf e}_x$ for $i=1$ and $N/2+1$ at $y=0$ is not applied in this model, although the squeezing force $-{\bf f}_i^{r}$ is applied for any $i$th particle. In this problem, the shear deformation plays an important role. 

\begin{figure}[t]
\begin{center}
\includegraphics[height=4.5cm]{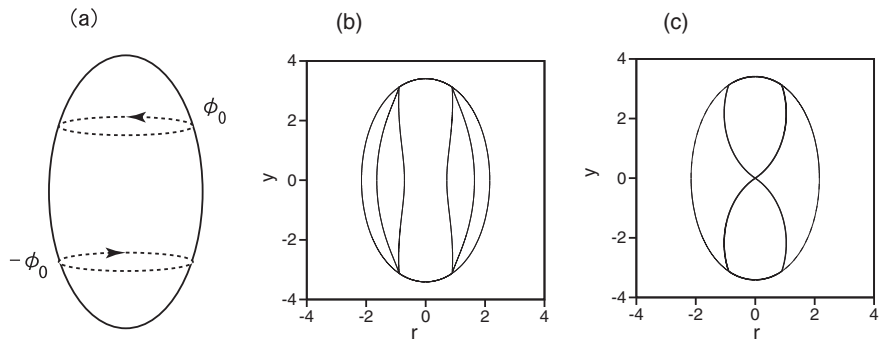}
\end{center}
\caption{(a) Twist of a spheroidal membrane. (b) Profiles of $(r_i,y_i)$ at $\phi_0=0,1.26$ and $2.06$ for $P=0.01,N=1000,K=64$, and $K^{\prime\prime}=4K\sin^2(\pi/N)$. (c) Profiles of $(r_i,y_i)$ at $\phi_0=0$ (thin curve) and $2.14$ (thick curve) for $P=0.01,N=1000,K=64$, and $K^{\prime\prime}=4K\sin^2(\pi/N)$.}
\label{f5}
\end{figure}
\begin{figure}
\begin{center}
\includegraphics[height=4.5cm]{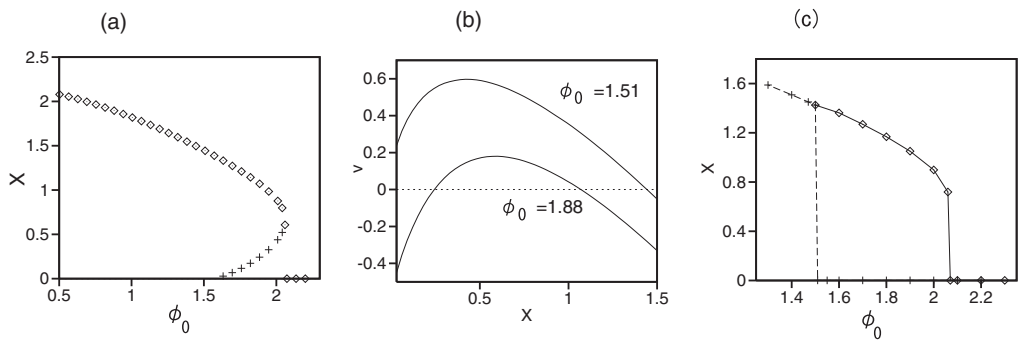}
\end{center}
\caption{(a) Radius $X$ at $i=1$ and $N/2+1$ as a function of $\phi_0$ for $P=0.01,N=1000,K=64$, and $K^{\prime\prime}=4K\sin^2(\pi/N)$. (b) Relationship of $v=dx_1/dt$ and $X$ for the fixed boundary conditions $(x_1,y_1,z_1)=(X,0,0)$ at $F=1.51$ and 1.88 for $P=0.01,N=1000,K=64$, and $K^{\prime\prime}=4K\sin^2(\pi/N)$. (c) Hysteresis loop of $X$ as a function of $\phi_0$. 
}
\label{f6}
\end{figure}
\begin{figure}
\begin{center}
\includegraphics[height=4.5cm]{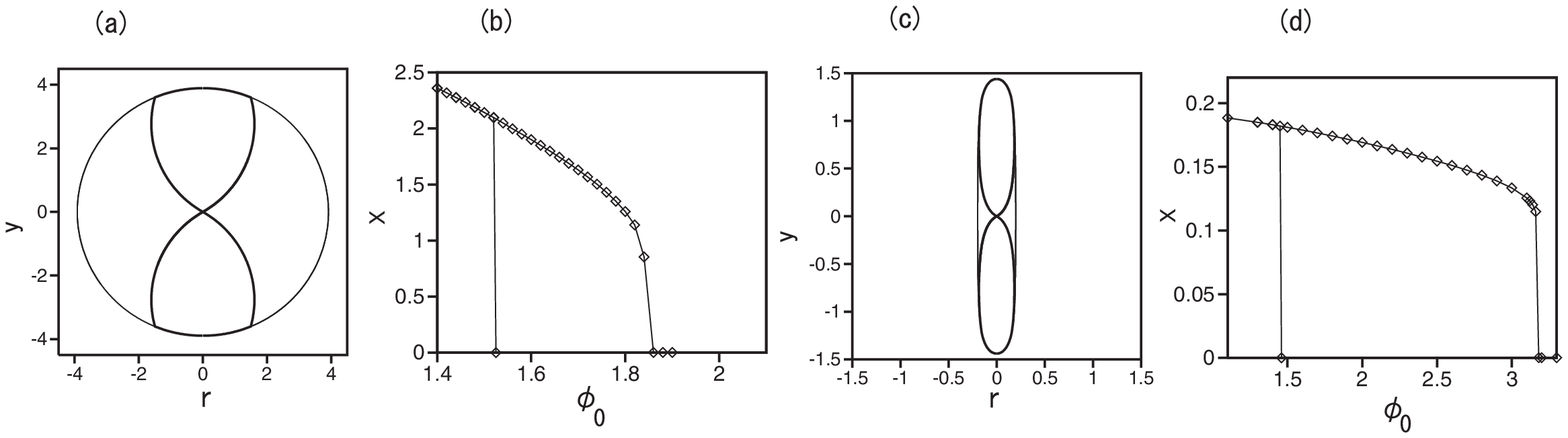}
\end{center}
\caption{(a) Profiles of $(r_i,y_i)$ at $\phi_0=0$ (thin curve) and $1.88$ (thick curve) for $P=0.01,N=1000,K=64$, and $K^{\prime}=0$. (b) Radius $X$ at $i=1$ and $N/2+1$ as a function of $\phi_0$ for $P=0.01,N=1000,K=64$, and $K^{\prime\prime}=0$. (c) Profiles of $(r_i,y_i)$ at $\phi_0=0$ (thin curve) and $3.2$ (thick curve) for $P=0.01,N=1000,K=64$, and $K^{\prime}=1280$. (d) Radius $X$ at $i=1$ and $N/2+1$ as a function of $\phi_0$ for $P=0.01,N=1000,K=64$, and $K^{\prime}=1280$. 
}
\label{f7}
\end{figure}
Figure 5(b) shows the profiles of $(r_i,y_i)$ where $r_i=\sqrt{x_i^2+z_i^2}$ at $\phi_0=0,1.26$, and 2.06 for $N=1000,K=64,K^{\prime\prime}=4K\sin^2(\pi/N)$ ($K^{\prime}=K$), and $P=0.01$. The profile of $(r_i,y_i)$ is interpreted as a projection of the two-dimensional surface into the plane of $z=0$. As $\phi_0$ is increased, the spheroidal membrane is wrung; thus, it becomes thinner owing to the twisting boundary conditions. The radius $X=r_i$ at the center $i=1$ and $N/2+1$ becomes zero at the critical value $\phi_{0c}$. This is because $r_i$'s for $i=1$ and $N/2+1$ are assumed to remain 0 by a repulsive force of very short range among neighboring particles, when $r_1$ or $r_{N/2+1}$ decreases to 0, since the reconnection of membranes does not occur in balloon art. 
The critical value is 2.07. The profile of $(r_i,y_i)$ at $\phi_0=2.14$ is shown in Fig.~5(c). A neck structure of width 0 appears at $y=0$.  Figure 6(a) shows the radius $X$ at $i=1$ and $N/2+1$ as a function of $\phi_0$. A discontinuous transition appears at $\phi_0\sim 2.07$. 
To find unstable stationary solutions, we have performed numerical simulations of Eq.~(8) under the fixed boundary conditions $x_1=-x_{N/2+1}=X,y_1=y_{N/2+1}=0$, and $z_1=z_{N/2+1}=0$. Figure 6(b) shows $v=dx_1/dt$ as a function of $X$ at $\phi_0=1.51$ and $1.88$. There are two stationary solutions satisfying $v(X)=0$ at $\phi_0=1.88$. The right one is the stable stationary solution and the left one is the unstable stationary solution. On the other hand, there is only one stable solution at $\phi_0=1.51$. The values of $X$ of the unstable solution are plotted by plus marks in Fig.~6(a). Figure 6(a) shows that the necking instability is a subcritical bifurcation.   
There is hysteresis and the necking structure is maintained stably for $1.515<\phi_0<2.06$, when $\phi_0$ is decreased from its upper critical value. 
Figure 6(c) shows a numerical result of the hysteresis loop, in which $\phi_0$ is stepwise increased from 1.5 to 2.3 and then decreased from 2.3 to 1.3. The rhombi denote the relationship of $\phi_0$ and $X$ in the increasing phase, and plus marks  denote that in the decreasing phase.  

We have further investigated the hysteresis by changing the parameter $K^{\prime}$ in $K^{\prime\prime}=4K^{\prime}\sin^2(\pi/N)$ with $N=1000$. Figure 7(a) shows the profiles of $(r_i,y_i)$ at $\phi_0=0$ and $1.88$  for $P=0.01,N=1000,K=64$, and $K^{\prime}=0$. When $K^{\prime}=0$, the profile of $(r_i,y_i)$ is circular at $\phi_0=0$. Figure 7(b) shows the relationship of $X=x_1=-x_{N/2+1}$ and $\phi_0$ for $K^{\prime}=0$. The upper critical value is 1.85 and the lower critical value is 1.525. Figure 7(c) shows the profiles  $(r_i,y_i)$ at $\phi_0=0$ and $3.2$  for $P=0.01,N=1000,K=64$, and $K^{\prime}=1280$. 
Owing to the strong squeezing force by the virtual ring chain, a cylindrical membrane appears at $\phi_0=0$. A neck structure appears at $\phi=3.2$ by the instability. Figure 7(d) shows  the relationship of $X=x_1=-x_{N/2+1}$ and $\phi$ for $K^{\prime}=1280$. The upper critical value is 3.17 and the lower critical value is 1.455. The parameter range of hysteresis increases with $K^{\prime}$, in particular, the upper critical value increases with $K^{\prime}$.   

In balloon art, a long and slender balloon is used. Hysteresis is important for  balloon art, in that the neck structures are stably maintained even if the twisting force is weakened. In our simple model Eq.~(8), the size and  shape of the stationary state are determined by the ratios of $P$, $K$, and $K^{\prime}$. That is, $P/K$ and $K^{\prime}/K$ are the control parameters. Furthermore, by the rescaling of space, i.e., ${\bf r}^{\prime}=\alpha {\bf r}$, the parameter $P/K$ can be set to a certain constant value; therefore, only the essential parameter is $K^{\prime}/K$. At $\phi_0=0$, the shape is spherical for $K^{\prime}/K=0$, and it becomes cylindrical as $K^{\prime}/K$ is increased.   
The parameter range of hysteresis is also determined by the ratio $K^{\prime}/K$.  We have performed numerical simulations of Eq.~(8) with the nondimensional parameters  $P$, $K$, and $K^{\prime}$; however, the angle $\phi_0$ has the unit of radian. For example, the result shown in Fig.~7(d) implies that a long and slender membrane exhibits the necking instability when the twist angle $\phi_0$  is increased to $3.17$ or $182^{\circ}$, and the necked state is maintained until the twist angle is decreased to $1.455$ or $83^{\circ}$. 

\section{Conclusion}
We have proposed a simple ring-chain model to simulate deformation in swollen membranes. We found a splitting instability caused by ring constriction. 
It is interpreted as one of the simplest models of cell splitting, in which  ring constriction causes such splitting. In  biological cytokinesis, a ring of microfilaments composed of actin is formed under the cell membrane. Myosin works as a motor protein and induces sliding motion among actin filaments, which leads to the ring constriction.  Although the detailed biophysical mechanism is not yet completely understood,  bending energy would be important in  cell division, because the biological membrane is a lipid bilayer membrane.  Our model can be extended to a bilayer membrane model composed of inner and outer ring chains to incorporate the effect of bending energy; however, this is left for future study.
In our model, the bifurcation can be subcritical or supercritical, depending on the conditions. In the simplest circular model, the bifurcation is subcritical and the critical value for remerging can be evaluated.  More detailed mathematical analysis might be possible in our simple model, because the model is very simple.

We have further investigated the twist deformation of a spheroidal membrane using the same one-dimensional model, and found a necking instability with hysteresis. The necking instability with hysteresis might be important in balloon art, in that created shapes are maintained even if the twisting forces are weakened. 
We have found that the parameter range of hysteresis increases with $K^{\prime}/K$, that is, the bistable parameter range is wider for a long and slender membrane. Some experiments using balloons or more suitable swollen membranes would be possible for checking and improving our simple model.  

\end{document}